\documentclass[11pt,a4paper]{article}

\usepackage[utf8]{inputenc}
\usepackage[T1]{fontenc}
\usepackage{amsmath}
\usepackage{amssymb}
\usepackage{graphicx}
\usepackage{hyperref}
\usepackage{geometry}
\usepackage{setspace}
\usepackage{authblk}
\usepackage{booktabs}
\usepackage{caption}
\usepackage{subcaption}
\usepackage{xcolor}
\usepackage[numbers,sort&compress]{natbib}
\usepackage{titlesec}
\usepackage{enumitem}
\usepackage{float}
\usepackage{array}
\usepackage{multirow}
\usepackage{url}
\usepackage{lineno}

\hypersetup{
  colorlinks=true,
  linkcolor=blue,
  citecolor=blue,
  urlcolor=blue
}

\titleformat{\section}{\large\bfseries}{\thesection.}{0.5em}{}
\titleformat{\subsection}{\normalsize\bfseries}{\thesubsection.}{0.5em}{}
\titleformat{\subsubsection}{\normalsize\itshape}{\thesubsubsection.}{0.5em}{}

\title{%
\rule{\textwidth}{0.5pt}\\[0.5ex]
\textbf{AAVGen: Precision Engineering of Adeno-associated Viral Capsids for Renal Selective Targeting}\\[0.5ex]
\rule{\textwidth}{0.5pt}
}

\author{%
  \large Mohammadreza Ghaffarzadeh-Esfahani$^{1}$,\;
  Yousof Gheisari$^{1,2,\dagger}$\\
  \small\itshape
  $^{1}$Regenerative Medicine Research Center, Isfahan University of Medical Sciences, Isfahan, Iran\\
  $^{2}$Department of Genetics and Molecular Biology, Isfahan University of Medical Sciences, Isfahan, Iran\\
  \small\normalfont
  $^{\dagger}$Correspondence: \href{mailto:ygheisari@med.mui.ac.ir}{ygheisari@med.mui.ac.ir},\;
  Tel/Fax: +98-3136687087
}
\date{\rule{\textwidth}{0.4pt}}

\begin{document}

% Abstract
\maketitle
\begin{abstract}
Adeno-associated viruses (AAVs) are promising vectors for gene therapy, but their native serotypes face limitations in tissue tropism, immune evasion, and production efficiency. Engineering capsids to overcome these hurdles is challenging due to the vast sequence space and the difficulty of simultaneously optimizing multiple functional properties. The complexity also adds when it comes to the kidney, which presents unique anatomical barriers and cellular targets that require precise and efficient vector engineering. Here, we present AAVGen, a generative artificial intelligence framework for de novo design of AAV capsids with enhanced multi-trait profiles. AAVGen integrates a protein language model (PLM) with supervised fine-tuning (SFT) and a reinforcement learning technique termed Group Sequence Policy Optimization (GSPO). The model is guided by a composite reward signal derived from three ESM-2-based regression predictors, each trained to predict a key property: production fitness, kidney tropism, and thermostability. Our results demonstrate that AAVGen produces a diverse library of novel VP1 protein sequences. In silico validations revealed that the majority of the generated variants have superior performance across all three employed indices, indicating successful multi-objective optimization. Furthermore, structural analysis via AlphaFold3 confirms that the generated sequences preserve the canonical capsid folding despite sequence diversification. AAVGen establishes a foundation for data-driven viral vector engineering, accelerating the development of next-generation AAV vectors with tailored functional characteristics.
\vspace{0.8cm}


\end{abstract}

\section{Introduction}

Gene therapy has emerged as a revolutionary approach for treating inherited and acquired diseases, with adeno-associated viruses (AAVs) standing as one of the most widely adopted vectors in clinical applications\cite{Yin2025}. These AAVs are highly valued as delivery vectors due to their non-pathogenic nature, long-term gene expression, and ability to infect different cell types\cite{Suarez2025}. However, wild-type (WT) AAV serotypes have limitations such as restricted tissue tropism, immune recognition, and variable transduction efficiency\cite{Pupo2022}. Engineered AAV capsids can evade pre-existing neutralizing antibodies\cite{Barnes2019}, penetrate physiological barriers such as the blood brain barrier\cite{Huang2023}, and achieve targeted delivery of therapeutic genes to desired tissues, including the liver\cite{Rodriguez2021}, muscle\cite{McGowan2025}, and the central nervous system\cite{Huang2023b}. Notably, the kidney represents a challenging target for AAV-mediated gene therapy, given its critical role in metabolic homeostasis and genetic disorders\cite{Issa2023}. The unique anatomical features of the kidney, including the glomerular filtration barrier and heterogeneous cellular populations, pose distinct challenges for efficient viral transduction\cite{Finch2023}. Existing AAV serotypes demonstrate limited kidney tropism and variable transduction efficiency across renal cell types, underscoring the need for precision-engineered vectors with enhanced renal selectivity\cite{Wu2024}.

Several techniques have emerged to engineer AAV capsids, each leveraging distinct principles to overcome the constraints of native serotypes. One foundational approach involves harnessing natural variants\cite{Lopez2024}, where sequences from diverse AAV isolates such as AAV2, AAV8, or AAV9 are shuffled or combined to incorporate advantageous traits, as exemplified by the creation of Anc80L65 for improved central nervous system transduction\cite{Hudry2018}. Subsequently, rational design relies on modification of the capsid using high-resolution structural insights to introduce targeted mutations, such as altering surface loops to modify receptor binding or inserting peptides for tissue-specific targeting, thereby enabling precise control over vector properties without exhaustive screening\cite{Lee2018}. Alternatively, directed evolution is used to generate large mutant libraries through error-prone PCR or DNA shuffling, followed by iterative selection in cell cultures or animal models for desired phenotypes\cite{Tabebordbar2021}. More recently, artificial intelligence-based approaches have accelerated capsid design by training predictive models on sequence-function datasets to predict the viability of generated capsids\cite{Tan2025}. While these methods have yielded clinically promising capsids such as those advancing in clinical trials for neuromuscular disorders\cite{Spencer2025}, they face important challenges like simultaneously optimizing multiple traits, such as production fitness, tissue tropism, and thermostability.

In this study, we have developed AAVGen, a generative protein language model (PLM) that integrates supervised fine-tuning (SFT)\cite{Ouyang2022} with reinforcement learning via Group Sequence Policy Optimization (GSPO)\cite{Zheng2025} to design functionally optimized AAV capsids for kidney tropism. By coupling large-scale PLMs with data-driven reward functions derived from experimentally validated assays, AAVGen learns the complex sequence--function relationships underlying AAV capsid performance. AAVGen is trained using three regression models derived from ESM-2 that estimate production fitness, renal tropism, and thermostability, and these predictions are integrated into a composite multi-objective reward function to steer the reinforcement learning process. This framework enables the de novo generation of diverse, biologically plausible AAV capsid variants with enhanced functional properties, while preserving the structural integrity of the WT scaffold. Together, these innovations establish AAVGen as a powerful platform for data-driven viral capsid engineering, offering a scalable and generalizable paradigm for multi-property optimization in synthetic capsid design.

% -----------------------------------------------------------------------
\section{Results}

\begin{figure}[!ht]
\centering
\includegraphics[width=0.8\textwidth]{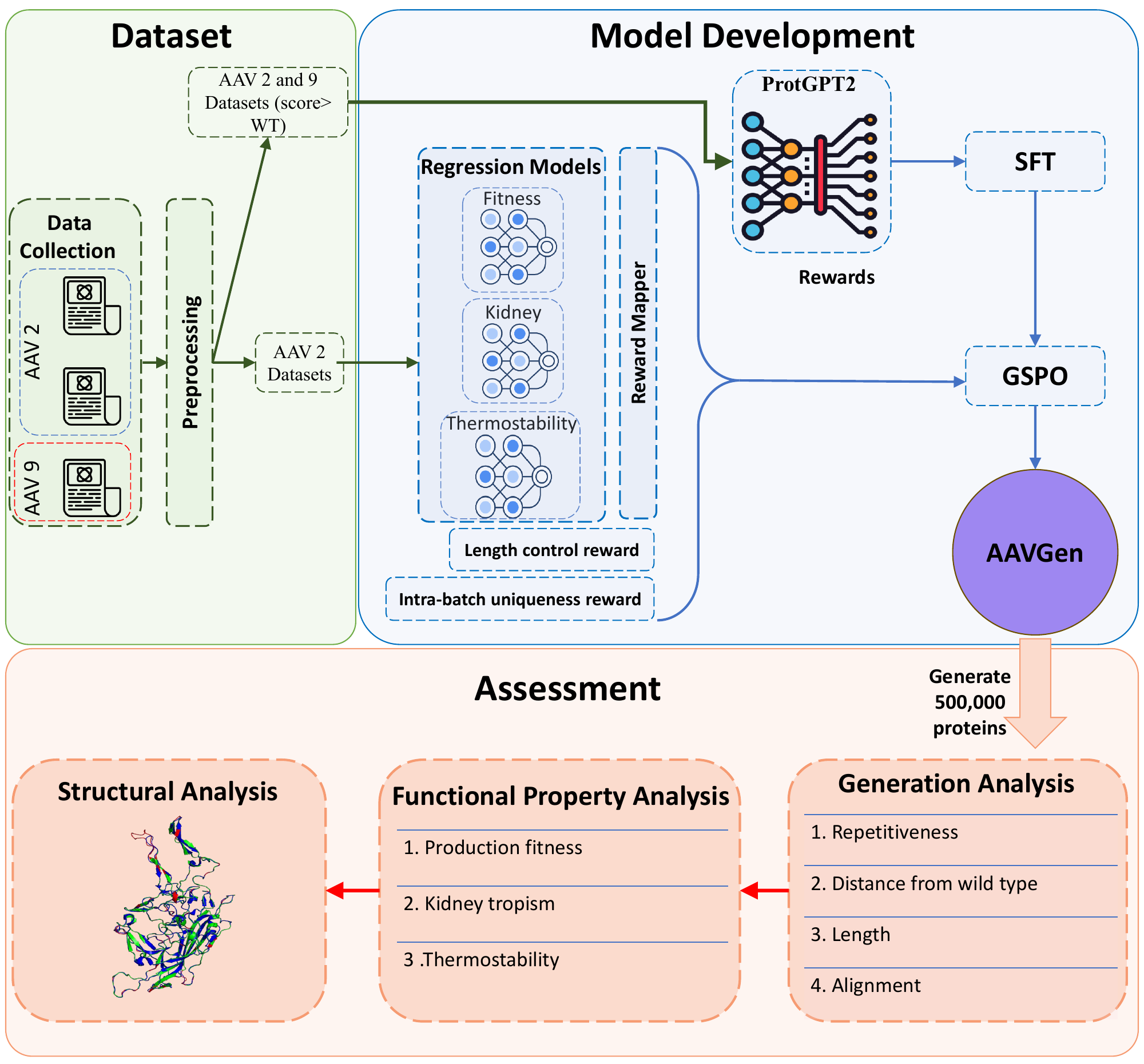} 
 \caption{\textbf{Development and assessment workflow of AAVGen.} The left upper panel illustrates the dataset curation process. The right upper panel details the model training phase, which includes supervised fine-tuning (SFT), custom reward modeling, and group sequence policy optimization (GSPO). The lower panel outlines the assessment procedure, comprising generation analysis, evaluation of production fitness, kidney tropism, thermostability, and structural analysis.}
\label{fig:1}
\end{figure}

To develop AAVGen, a generative PLM for designing AAV capsids, we used SFT along with reinforcement learning via GSPO. The process began by training three ESM-2-based regression models\cite{Lin2023} to serve as reward functions, predicting production fitness, kidney tropism, and thermostability from sequence. Production fitness reflects the capsid's packaging efficiency during expression, kidney tropism quantifies the vector's ability to transduce the kidney, and thermostability measures resistance to thermal degradation, critical for storage and in vivo delivery. The ProtGPT2 model\cite{Ferruz2022} was first fine-tuned on AAV2 and AAV9 VP1 datasets to learn foundational residue--residue relationships. GSPO was then applied to refine this model, using a composite reward from the regression predictors to promote the generation of capsids with improved multi-property profiles. Finally, we assessed the generated sequences based on their uniqueness, sequence length distribution, and alignment. We also used regression scoring and AlphaFold3 structural modeling to compare their predicted function and structure to WT AAV2. The schematic design of the study is illustrated in Figure~\ref{fig:1}.

\subsection{Regression models effectively predict production fitness, kidney tropism, and thermostability of AAV variants}

\begin{figure}[!ht]
\centering
\includegraphics[width=0.8\textwidth]{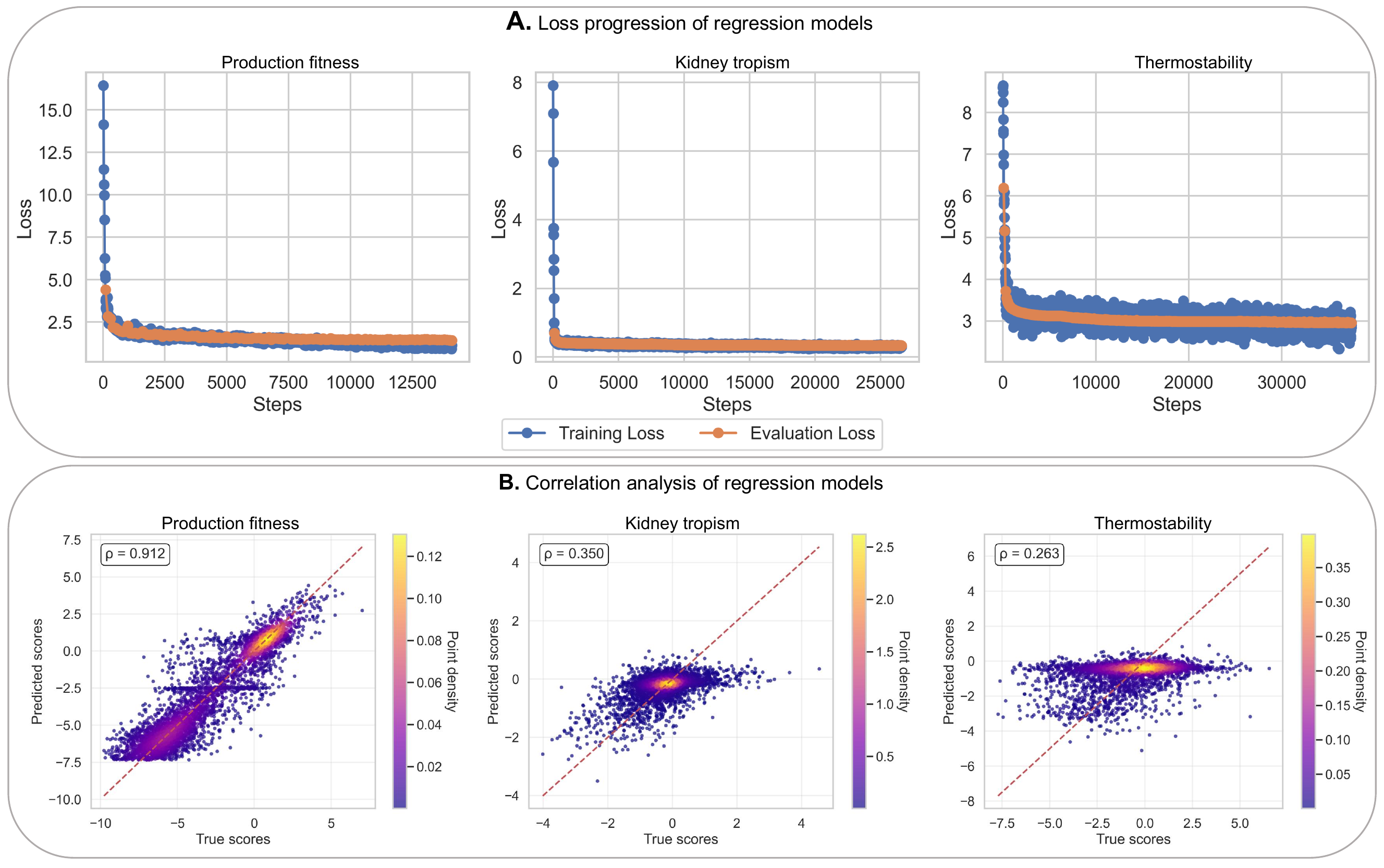} 
\caption{\textbf{Evaluation of regression models.} \textbf{(A)} Training loss progression for models predicting production fitness, kidney tropism, and thermostability. \textbf{(B)} Correlation between experimentally determined (true) and model-predicted scores for production fitness, kidney tropism, and thermostability.}
\label{fig:2}
\end{figure}

To develop AAVGen, we trained three regression models on the AAV2 dataset to predict production fitness, kidney tropism, and thermostability of generated capsid variants. Regression models were employed to capture the correlation between sequences and experimentally measured values. The ESM-2 model was first fine-tuned on the VP1 sequence--fitness dataset until the loss plateaued at approximately 14,100 training steps. Subsequently, the fitness model was further fine-tuned on the kidney tropism dataset to transfer knowledge from the fitness dataset and trained for 26,600 steps. Similarly, the fitness model was used to fine-tune a thermostability model on its respective dataset until convergence at 37,500 steps (Figure~\ref{fig:2}A, Supplementary files \href{https://github.com/mohammad-gh009/AAVGen/blob/main/assets/Supplementary/Supplementary_file_1.json}{1}, \href{https://github.com/mohammad-gh009/AAVGen/blob/main/assets/Supplementary/Supplementary_file_2.json}{2}, and \href{https://github.com/mohammad-gh009/AAVGen/blob/main/assets/Supplementary/Supplementary_file_3.json}{3}). Furthermore, the predictive performance of models was evaluated by measuring the correlation between predicted and actual values. The model trained for production fitness achieved a Spearman correlation coefficient of 0.91 with a p-value of $<10^{-n}$, demonstrating strong predictive performance. The kidney tropism model showed a moderate correlation (Spearman $\rho = 0.35$, p-value $<10^{-n}$), while the thermostability model exhibited a weaker but statistically significant correlation (Spearman $\rho = 0.26$, p-value $<10^{-n}$) (Figure~\ref{fig:2}B).

\subsection{Policy optimization with customized rewards produced enhanced AAV variants}

\begin{figure}[!ht]
\centering
\includegraphics[width=0.8\textwidth]{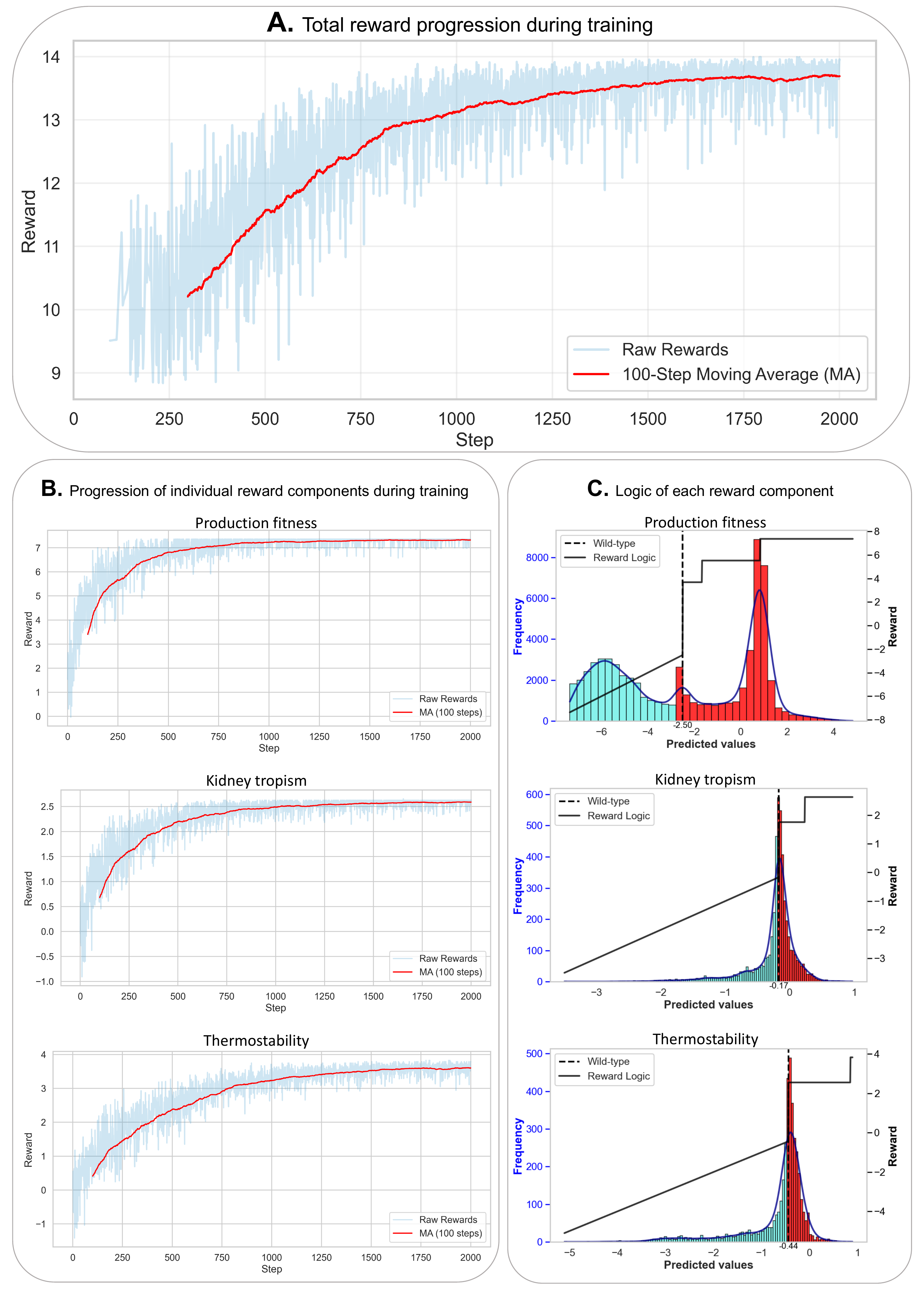} 
\caption{\textbf{Reward progression in training with group sequence policy optimization (GSPO).} \textbf{(A)} Total reward over the course of training. \textbf{(B)} Reward progression during training of AAVGen for reward functions that produce production fitness, kidney tropism, and thermostability. \textbf{(C)} Reward assignment logic for each objective, defined by the mean absolute error of model predictions on the validation set relative to the wild-type (WT) reference for production fitness, kidney tropism, and thermostability.}
\label{fig:3}
\end{figure}

To guide AAVGen training with the GSPO framework, we used three regression models trained before as reward functions to fine-tune a ProtGPT2-based model that had been previously fine-tuned on AAV2 and AAV9 datasets. Training proceeded until the total composite reward function plateaued, indicating convergence (Figure~\ref{fig:3}A, Supplementary file \href{https://github.com/mohammad-gh009/AAVGen/blob/main/assets/Supplementary/Supplementary_file_4.json}{4}). This composite reward consisted of three main components: production fitness, kidney tropism, and thermostability rewards (Figure~\ref{fig:3}B) plus two auxiliary rewards that stabilized training, including length control reward which prevents the model from collapsing into generating sequences identical or highly similar to the WT length and uniqueness reward which encourages the model to generate unique sequences in each training batch (Supplementary file \href{https://github.com/mohammad-gh009/AAVGen/blob/main/assets/Supplementary/Supplementary_file_5.docx}{5}).

To obtain a set of high-quality capsid variants with improved properties relative to the WT and to mitigate potential prediction errors from the regression models, we designed a reward logic mapper that translates the predicted scores from each of these three models into corresponding reward signals by categorizing scores exceeding the WT sequence (Figure~\ref{fig:3}C). Through this combination of primary and auxiliary rewards, along with the reward logic mapper, the model achieved a stable optimization state, as evidenced by each reward component reaching a plateau during training.

\subsection{AAVGen produces a diverse and biologically plausible capsid library}

\begin{figure}[!ht]
\centering
\includegraphics[width=0.8\textwidth]{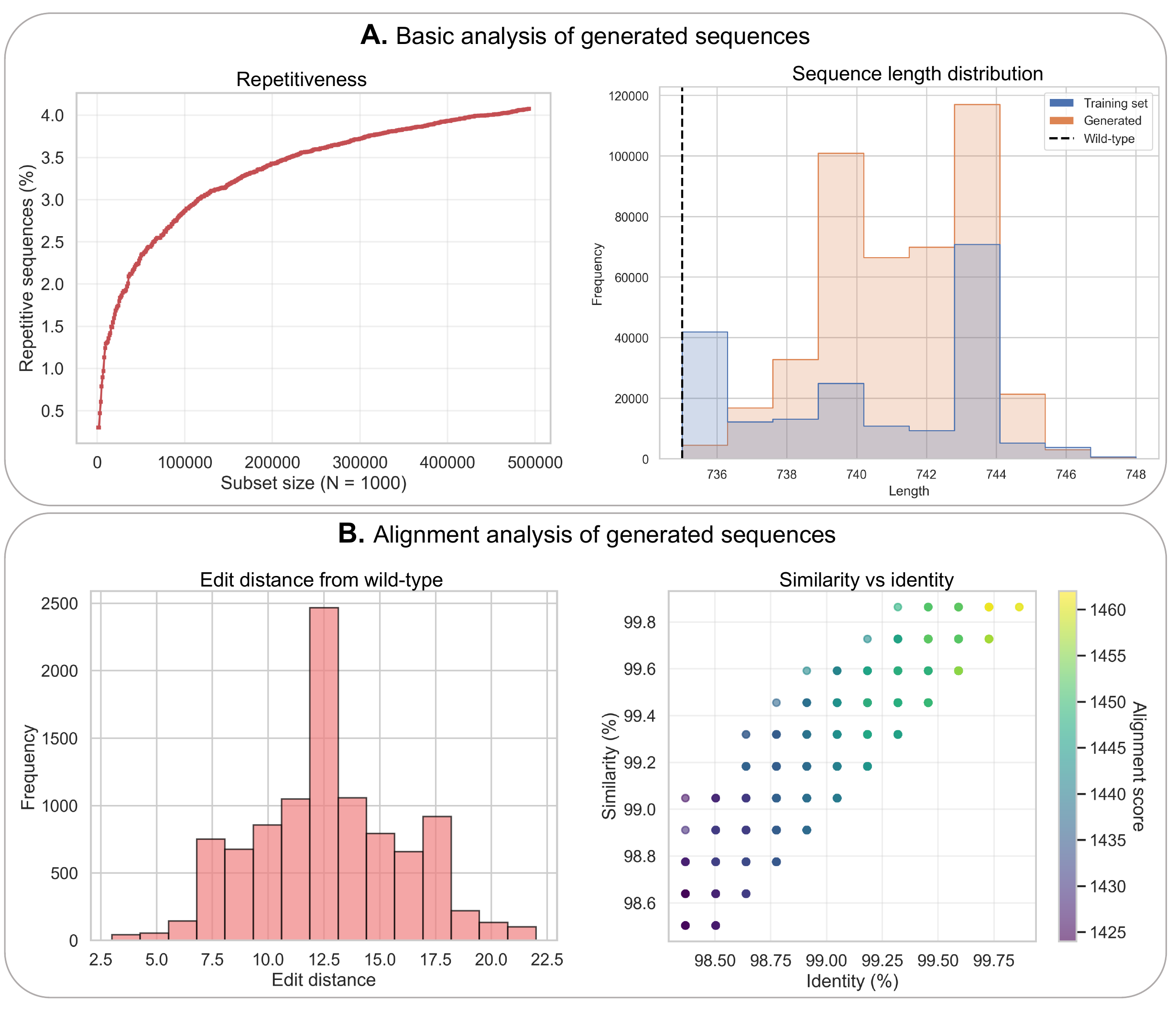} 
\caption{\textbf{Sequence diversity and alignment metrics of the AAVGen-generated library.} \textbf{(A)} Performance benchmarks of the generative model regarding sequence novelty. Left: Cumulative repetitiveness of generated sequences across increasing subset sizes ($N = 1{,}000$) for a library of 500,000 variants. Right: Distribution of sequence lengths for the training set (blue) and generated sequences (orange) relative to the wild-type (WT) AAV2 VP1 sequence (dashed line). \textbf{(B)} Alignment-based divergence of generated variants from the AAV2 WT reference. Left: Frequency distribution of edit distances from the WT sequence. The majority of synthetic variants contain between 10 and 15 mutations compared to the WT template. Right: Correlation between sequence similarity and identity percentages. Data points are colored by alignment score (ranging from 1425 to $>1460$), illustrating that the generated sequences maintain high conservation ($>98.5\%$ identity) while exploring a diverse landscape of substitution patterns.}
\label{fig:4}
\end{figure}

To evaluate the generative capability of AAVGen, we analyzed a set of 500,000 generated protein sequences across three key metrics. First, we measured the uniqueness of generation to understand how many repetitive sequences our model generates and to quantify the model's capacity to explore diverse regions of sequence. Second, we evaluated length distribution to confirm biologically realistic protein dimensions, as deviations from natural VP1 length could compromise capsid assembly and stability (Figure~\ref{fig:4}A). Finally, we assessed the fidelity to the WT AAV2 to measure the difference between the generated sequences and the WT (Figure~\ref{fig:4}B).

Our analysis of sequence uniqueness revealed that approximately 4\% of the generated set was repetitive. Furthermore, after removing duplicates, 1,787 sequences were exact matches to those in the training set. When compared to WT sequences, 230 generated sequences were identical to AAV2, while none matched the AAV9 WT, indicating a low rate of exact template replication and a successful bias toward generating novel variants. The length distribution of the generated sequences (median: 741; IQR: 740--743) closely matched that of the training set (median: 741; IQR: 737--743). The generated sequences showed high sequence similarity---the percentage of aligned positions with either identical or biochemically similar amino acids---(median: 99.32\%, IQR: 99.05--99.46\%) and identity---the percentage of aligned positions with identical amino acids---(median: 99.18\%, IQR: 98.91--99.32\%) to AAV2, a finding further supported by a high alignment score---a numerical value reflecting the overall quality of a sequence alignment---rewarding matches and penalizing mismatches and gaps (median: 1443.5; IQR: 1438.5--1448.5). We also used edit distance\cite{Berger2021}---the minimum number of single-amino-acid substitutions, insertions, or deletions required to convert one sequence into another---to quantify sequence divergence, finding a median difference of 13\% (IQR: 10--15\%) from the AAV2 WT. Collectively, these results confirm that AAVGen successfully generates a diverse and expansive library of novel capsid variants that retain the core structural and functional hallmarks of WT AAVs.

\subsection{AAVGen generates high-quality sequences in terms of production fitness, kidney tropism, and thermostability}

\begin{figure}[!ht]
\centering
\includegraphics[width=0.8\textwidth]{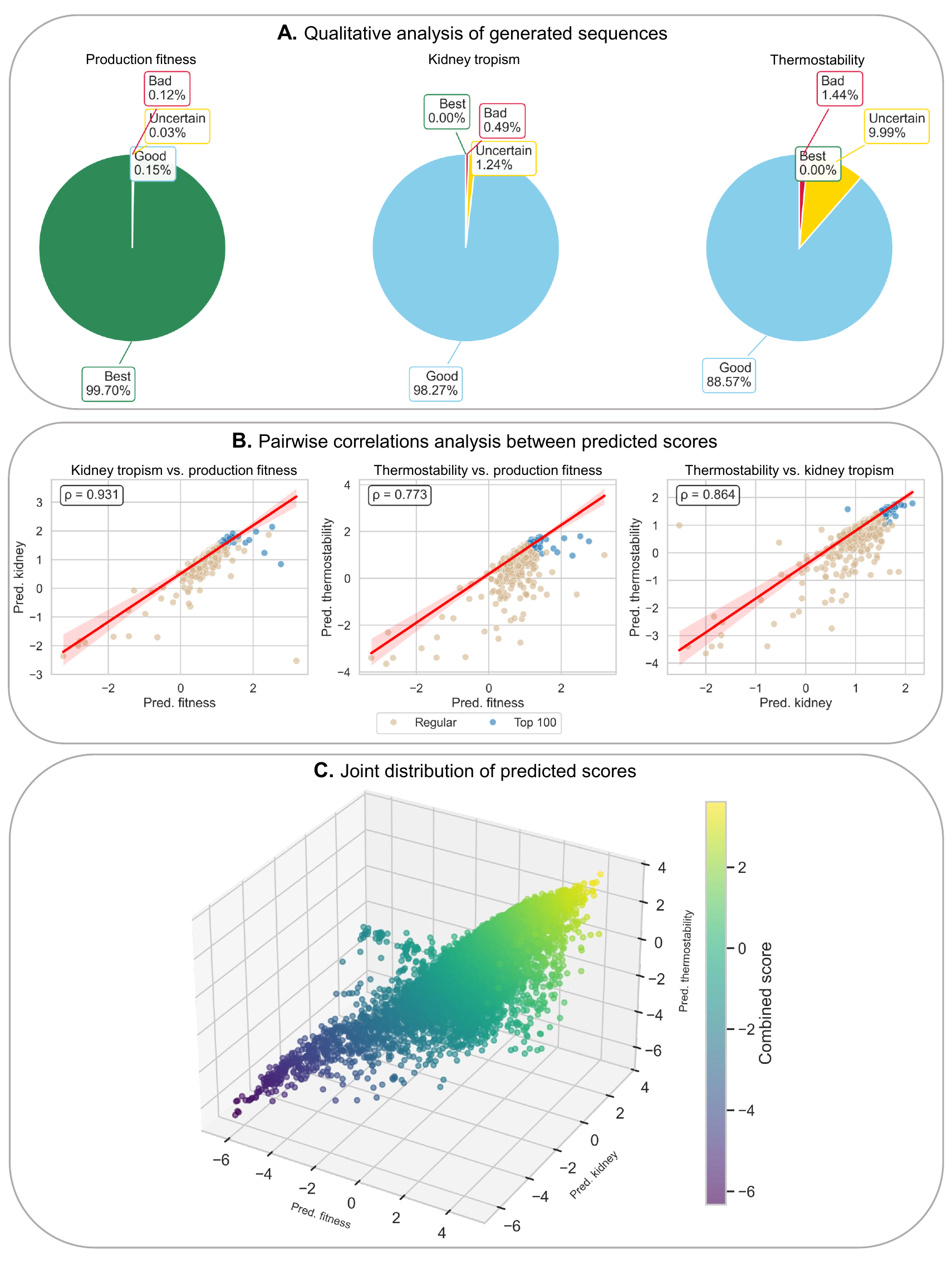}
 \caption{\textbf{Functional property analysis of generated AAVGen sequences.} \textbf{(A)} Qualitative classification of generated sequences into “Best”, “Good”, “Uncertain”, and “Bad” categories based on predicted production fitness, kidney tropism, and thermostability scores. \textbf{(B)} Pairwise correlation analyses between predicted production fitness, kidney tropism, and thermostability scores. \textbf{(C)} Joint three-dimensional distribution of predicted production fitness, kidney tropism, and thermostability scores, with each point colored according to the average of these three scores.}
\label{fig:5}
\end{figure}

To evaluate the functional quality of the generated capsids, we assessed the quality of 500,000 generated sequences. After filtering out repetitive, WT, and training-set duplicates, 436,765 sequences remained. Using regression models, sequences were categorized based on their predicted scores relative to WT and the model's mean absolute error (MAE) on the validation set margin: ``Best'' (exceeding WT $+$ 4 MAE), ``Good'' (between WT $+$ 1 and $+$4 MAE), ``Uncertain'' (between WT and WT $+$ 1 MAE), and ``Bad'' for sequences that have a lower than WT predicted score. The analysis demonstrates that AAVGen consistently produces sequences with high predicted performance across all three key properties (Figure~\ref{fig:5}A).

Regarding production fitness, the model exhibited exceptional performance in generating high-fitness sequences. The vast majority of records, 435,448 sequences (99.7\%), were classified as ``Best,'' while 669 (0.15\%) were ``Good'', 128 (0.03\%) and 559 (0.4\%) sequences were ``Uncertain'' and ``Bad'' respectively. Additionally, for kidney tropism, the output was predominantly composed of ``Good'' sequences (491,439; 98.27\%). A smaller proportion were ``Uncertain'' (5,416; 1.24\%), 2,155 (0.4\%) were ``Bad'', and one sequence achieved the ``Best'' classification. For thermostability, no sequences met the criteria for the ``Best'' category. Nonetheless, the model produced a large number of variants with improved predicted thermostability, with 386,844 sequences (88.57\%) classified as ``Good'', alongside smaller fractions categorized as ``Uncertain'' (43,626; 9.99\%) or ``Bad'' (6,295; 1.44\%).

Having confirmed that the majority of generated sequences were classified as ``Good'' or ``Best'' with raw predictions exceeding WT levels, we next investigated whether these desirable properties tend to co-occur within individual sequences. To address this, we computed Spearman correlations between the predicted scores for each pair of properties. We observed strong positive correlations across all optimized properties, indicating that improvements in one property are consistently associated with improvements in the others (Figure~\ref{fig:5}C). This finding demonstrates that AAVGen successfully generates sequences that jointly optimize for all three design objectives rather than trading one property for another (Figure~\ref{fig:5}D).

\subsection{AAVGen diversifies sequences while preserving the structural scaffold}

To further assess the structural robustness and diversity of the generated sequences, we randomly selected 500 proteins from the ``Good'' and ``Best'' classes among the 500,000 generated sequences. For each protein, five independent structures were predicted using AlphaFold3\cite{Abramson2024}. To establish a baseline, we randomly generated 250 VP1 sequences by identifying the variable regions of these 500 sequences relative to the WT through sequence alignment and inserting random amino acids matched to the length distribution of the 500 generated sequences. Each baseline sequence was folded using the same strategy. Structural similarity to the WT protein was quantified by calculating the root-mean-square deviation (RMSD). For each sequence, the median RMSD across the five predicted structures was considered as the representative value.

\begin{figure}[!ht]
\centering
\includegraphics[width=0.8\textwidth]{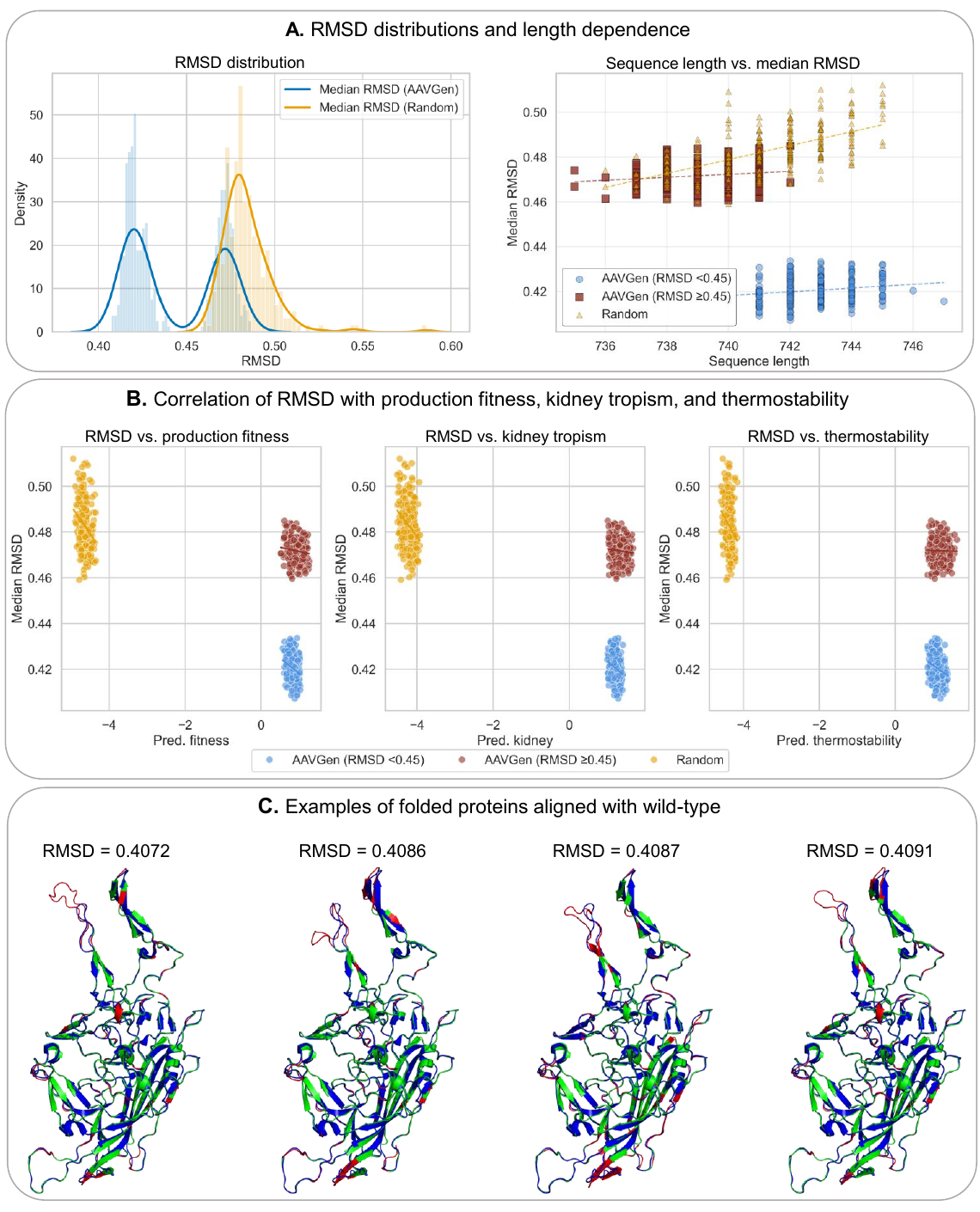} 
 \caption{\textbf{Structural analysis of the generated sequences.} \textbf{(A)} Left: Distribution of Root Mean Square Deviation (RMSD) for $n = 500$ sampled protein sequences from AAVGen and $n = 250$ randomly generated relative to the wild-type (WT) reference. For each sequence, the median RMSD was calculated across five independent AlphaFold3 structure predictions. Right: Relationship between sequence length and median RMSD, indicating a modest increase in RMSD with longer sequences, while AAVGen sequences consistently maintain lower RMSD compared with random sequences. \textbf{(B)} Correlation analysis between median RMSD and predicted functional attributes. \textbf{(C)} Representative structural alignments of the four generated sequences exhibiting the lowest median RMSD. Structures are superimposed on the WT model. Green regions denote high spatial conservation, where C$\alpha$ atoms lie within 0.5~\AA{} of the WT reference. Structural divergence is highlighted by red (non-conserved residues in the mutant) and blue (corresponding residues in the WT).}
\label{fig:6}
\end{figure}

As shown in Figure~\ref{fig:6}A, the RMSD distribution of AAVGen-generated sequences exhibited two distinct modes: one centered around 0.42~\AA{} and another around 0.47~\AA{}, compared with a median RMSD of 0.48~\AA{} for the randomly generated baseline sequences. While RMSD generally increases with sequence length across all groups, this relationship was substantially weaker for both AAVGen low and high RMSD subgroups (Spearman $\rho = 0.17$ and 0.20, respectively) compared to the random baseline (Spearman $\rho = 0.61$), suggesting that AAVGen's sequence diversity is not simply a byproduct of length variation.

Subsequently, we examined the relationship between RMSD and predicted production fitness, kidney tropism, and thermostability within each group (Figure~\ref{fig:6}B). In the lower-RMSD sub-group of AAVGen (RMSD $< 0.45$~\AA{}), we observed statistically significant negative Spearman correlations between RMSD and all three metrics: production fitness ($\rho = -0.22$), kidney tropism ($\rho = -0.23$), and thermostability ($\rho = -0.25$), indicating that sequences more structurally similar to the WT tend to score higher across functional predictions. In contrast, no significant correlations were detected in the higher-RMSD group (RMSD $\geq 0.45$~\AA{}), suggesting that beyond a structural deviation threshold, predicted functional properties decouple from structural similarity to the WT. Notably, both AAVGen subgroups vastly outperformed the random baseline across all functional metrics. In the lower-RMSD group, median predicted production fitness, kidney tropism, and thermostability were 0.81, 1.21, and 1.10, respectively, while the higher-RMSD group scored even more favorably at 0.92, 1.34, and 1.22. Both stand in sharp contrast to the random baseline, which yielded median values of $-4.65$, $-4.19$, and $-4.37$ for the same metrics. This consistent and substantial gap across all three dimensions confirms that AAVGen generates functionally plausible variants with meaningful biological properties, rather than sequences that happen to resemble the WT structure by chance. Finally, the four variants with the lowest RMSD values are visualized in Figure~\ref{fig:6}C.

% -----------------------------------------------------------------------
\section{Discussion}

In this study, we introduced AAVGen, a generative PLM for designing AAV capsids with enhanced functional properties. By integrating supervised fine-tuning of the ProtGPT2 model on a diverse corpus of AAV sequences with GSPO, we developed a model capable of generating novel AAV2 VP1 protein variants. Our results demonstrate that AAVGen successfully produces a vast and diverse library of capsid sequences that are highly novel yet retain strong structural similarity to WT AAV2. Importantly, the generated variants are predicted by specialized regression models to exhibit enhanced production fitness, kidney tropism, and thermostability, showcasing the model's ability to perform multi-trait optimization and generate high-quality, biologically plausible AAV candidates for further development.

Building on these results, we developed a generalizable framework for training AAVGen for high-quality, versatile AAV sequence generation, enabling effective generation of protein sequences through a comprehensive understanding of the VP1 capsid protein. To achieve this, we constructed regression models capable of processing and scoring VP1 sequences of variable lengths, unlike prior studies such as Bryant et al.\cite{Bryant2021}, who employed different neural networks, and Eid et al.\cite{Eid2024}, who used LSTM-based regression limited to fixed-length inputs. Our approach thus allows for the generation of proteins with variable lengths while maintaining sequence quality. This was accomplished by fine-tuning the ESM-2 model to interpret the entire sequence context and perform scoring directly. In contrast to prior methods such as CAP-PLM\cite{Wu2025}, which relied on extracted embeddings of ESM-2 and one-hot encodings for regression training, our method fine-tunes the full ESM-2 model jointly with the embedding component. This approach outperformed the earlier technique, achieving superior predictive performance. Nevertheless, while regression and classification models applied to fitness datasets have shown promising results, the development of a multifaceted and multi-objective AAV design framework remains a major challenge. Zheng et al.\cite{ZhengH2024} attempted to address this by introducing the ALICE system, which leverages contrastive and reinforcement learning to guide AAV capsid generation. However, ALICE's reinforcement learning phase focuses solely on CNS tropism and fails to account for other attributes such as thermostability. Additionally, the reliance of their method on MSA considerably slows training, particularly when scaling to larger datasets.

An additional novel aspect of our study is the use of multiple AAV serotypes (AAV2 and AAV9) to fine-tune our base model. Liu et al.\cite{Liu2024} have shown that introducing the same mutations from AAV2 into the AAV9 sequence can enhance its fitness, suggesting that different serotypes share meaningful functional and structural relationships. By leveraging sequences from multiple serotypes during model training, we were able to capture these shared features, improving the generalizability and effectiveness of AAVGen. This approach, alongside multiple-objective training, enabled a more capable model that addresses previous challenges while enabling comprehensive and scalable AAV design.

Based on these methodological foundations, AAVGen effectively learns and exploits the complex sequence--function relationships underlying AAV capsid performance to design variants with superior multi-trait characteristics. The regression models trained on experimental datasets captured distinct yet interrelated aspects of AAV biology, as reflected by their strong predictive correlations, particularly for production fitness. Integration of these models as reward functions in the GSPO framework enabled AAVGen to perform multi-objective optimization, simultaneously improving fitness, kidney tropism, and thermostability. The generated library of 500,000 sequences exhibited high novelty and diversity while maintaining structural fidelity to WT AAV2. Notably, the vast majority of generated variants displayed predicted fitness surpassing that of the WT, and substantial subsets showed concurrent enhancement in kidney tropism and thermostability. The strong positive correlations among the optimized properties indicate that AAVGen captures shared evolutionary and biophysical constraints governing AAV capsid stability and tropism, suggesting that reinforcement learning guided by multi-property rewards can steer generative models toward biologically coherent and functionally balanced solutions. Collectively, these findings validate the capacity of AAVGen to generate high-quality, biologically plausible AAV capsid sequences with improved multi-trait performance.

Complementing these multi-trait functional improvements, structural analysis of the AAVGen's generated sequences uncovered an intriguing aspect of the optimization landscape. Specifically, the RMSD values of the generated variants relative to the WT exhibited a bimodal distribution. This pattern likely arises from the design of the reward functions and the nature of multi-objective optimization performed by the model. Rather than converging on a single structural optimum, AAVGen appears to have learned two primary modes of high-performing solutions, one closely resembling the WT conformation and another representing a distinct, yet still viable, structural variant. Both modes yield sequences with substantially improved predicted functional properties compared to random baseline sequences, underscoring the model's ability to explore biologically meaningful regions of sequence space while maintaining overall capsid integrity.

Although our study yielded promising results, it has several limitations. Foremost, the generated sequence was not experimentally validated in vitro, which precludes definitive conclusions regarding its functional performance. Additionally, training regression models for kidney tropism and thermostability was hindered by the scarcity of high-quality data, resulting in limited predictive power and higher MAE on the validation set. Future studies could benefit from incorporating wet-lab experiments and employing strategies such as those described by Jiang et al.\cite{Jiang2025}, which involve modifying non-surface amino acids to generate viable capsids. Such approaches could expand the dataset size and improve model performance.

In summary, this study demonstrates that AAVGen represents a significant advancement in AAV capsid design, combining generative modeling with multi-objective optimization to produce highly diverse, novel, and biologically plausible variants. By leveraging sequence--function relationships through fine-tuned regression models and GSPO-guided generation, AAVGen effectively enhances multiple functional traits while maintaining structural fidelity. Although experimental validation and expanded datasets remain necessary to fully confirm its predictions, our findings establish a scalable and versatile framework for AAV engineering, paving the way for the development of next-generation gene therapy vectors with tailored properties.

% -----------------------------------------------------------------------
\section{Methods}

\subsection{Data collection}

To develop AAVGen, we integrated AAV capsid fitness data from three independent studies. A comprehensive fitness landscape for AAV2 was established by Ogden et al.\cite{Ogden2019} through deep mutational scanning, which included libraries of 31,579 VP1 sequences assessed for production fitness, 24,984 sequences assessed for kidney tropism, and 30,889 sequences evaluated for thermostability. To expand the mutational scope, we incorporated data from Bryant et al.'s study\cite{Bryant2021}, which focused on a specific region (residues 561--588) of the AAV2 VP1 capsid. This study provided a large-scale multiple-mutation dataset comprising 296,896 sequences with associated fitness scores. Finally, to include data on a different serotype and additional functional attributes, we utilized the AAV9 dataset from Eid et al.\cite{Eid2024}. This dataset comprises 100,000 sequences characterized for production fitness, along with key functional outcomes, including binding/transduction in HEPG2 and THLE-2 cell lines and liver biodistribution in a mouse model. The combination of these datasets provided a robust and multifaceted foundation for training AAVGen.

\subsection{Data pre-processing}

To construct a unified training dataset from the three independent studies, we implemented a comprehensive processing pipeline. The primary challenge was harmonizing diverse data formats, mutation types (insertions, substitutions, and deletions), and experimental assays into a consistent structure suitable for model input. Our strategy involved reconstructing the full-length VP1 amino acid sequence for each variant, normalizing fitness scores to a common scale, and removing low-quality or ambiguous data points. The following sections describe the specific processing steps.

\subsubsection{Data preparation for regression models}

We fine-tuned AAVGen to generate high-quality AAV2 capsids by developing three distinct regression models to predict production fitness, kidney tropism, and thermostability of the capsid. Two studies that provided the AAV2 datasets were used to fine-tune regression models. These models were constructed by fine-tuning the ESM-2 PLM to enable estimation of functional properties from amino acid sequences. The resulting regression models served as reward functions within the GSPO framework, guiding the generation process toward variants with superior functional properties.

\paragraph{Production fitness.}
The production fitness dataset was compiled by processing data from two studies: Bryant et al. and Ogden et al. The dataset from Bryant et al.'s study provided amino acids and corresponding fitness scores for VP1 positions 561--588, which were normalized to the WT score and $\log_2$-transformed. From the Ogden et al. study, production fitness was derived from a packaging dataset. We obtained deep mutational scanning read counts for AAV2 variants from this source and excluded technical replicates with low coverage. Selection values were then calculated using variant frequencies from plasmid and viral pools, normalized to the WT score, and $\log_2$-transformed. The insertion, substitution, and deletion libraries were processed separately. For each variant, the full-length AAV2 sequence was reconstructed by applying the respective mutation. Finally, the datasets from both studies were merged into a single variant-level table containing the reconstructed sequence and its normalized production fitness score.

\paragraph{Kidney tropism.}
To process the kidney tropism dataset, the packing dataset was loaded from the Ogden et al. study, and technical replicates with poor coverage were excluded. Mouse variant counts were further filtered to remove low-abundance variants ($<10$ reads) and extremely high-abundance variants ($>31{,}000$ reads) to reduce noise from sequencing artifacts. Selection values were computed as the ratio of variant frequencies in mouse tissue samples to their corresponding frequencies in packaged viral pools, normalized to WT variants. Furthermore, replicate measurements were aggregated, and median selection coefficients were calculated. Selection scores were $\log_2$-transformed, and non-finite values were removed. For reconstruction of VP1, insertion, substitution, and deletion libraries were processed separately. Full-length AAV2 capsid sequences were regenerated for each variant by applying the corresponding mutation to the reference sequence. Insertion, substitution, and deletion datasets were merged, yielding a final dataset that included the reconstructed sequence and the normalized kidney tropism score.

\paragraph{Thermostability.}
The thermostability dataset was constructed using raw AAV2 variant counts obtained from the packaging datasets reported by the Ogden et al. study. This assay measured capsid thermostability by incubating the library at different temperatures and then digesting any exposed genomes. For data processing, technical replicates with low coverage were first excluded. We then calculated selection values by normalizing variant frequencies in the thermostability-selected pools against their corresponding packaged viral frequencies. The resulting thermostability scores were averaged across replicates and $\log_2$-transformed. The insertion, substitution, and deletion libraries were processed separately. Finally, the datasets were merged into a unified table containing the reconstructed sequence and normalized thermostability score, from which any variants with missing or non-finite scores were removed.

Each dataset was subsequently divided into training and validation subsets using a stratified sampling method. Targets were first grouped into 10 quantiles to maintain their distribution across both subsets. This process ensured that the datasets were properly prepared for training the regression models.

\subsubsection{Data preparation for SFT}

To develop a generalizable generative model capable of understanding the residue-residue relationships within AAV capsids, we fine-tuned ProtGPT2 on a diverse collection of VP1 sequences. This included integrating variants from multiple serotypes, specifically AAV2 and AAV9, to expose the model to a broader spectrum of viable amino acid combinations. This foundational step allowed the model to learn the underlying grammar of functional AAV capsids.

For supervised fine-tuning of the model, we integrated a comprehensive collection of VP1 sequences from the AAV2 (Bryant et al., Ogden et al.) and AAV9 (Eid et al.) studies to teach the model the residue-residue relationships across serotypes. The AAV9 dataset, which characterizes variants based on production fitness, binding/transduction in HEPG2 and THLE-2 cell lines, and liver biodistribution in mice, required specific processing. We generated full-length AAV9 VP1 sequences by using the WT serotype of AAV9 capsid as a reference and inserting each variant peptide at the specified site between residues 588 and 589. To select high-performing variants, we included all sequences with fitness scores exceeding that of the WT. These selected sequences were then deduplicated to create the final dataset.

The final consolidated corpus contained 192,199 non-redundant VP1 sequences, each annotated with its serotype (AAV2 or AAV9). To ensure compatibility with the ProtGPT2 model architecture, sequences were formatted into a FASTA-like structure with amino acids grouped into blocks of 60 characters. An end-of-sequence (EOS) token was appended to the start and end of each formatted sequence. This comprehensive dataset was then partitioned into training (80\%) and validation (20\%) subsets using a fixed random seed, with stratification by serotype to maintain a balanced representation of both AAV types.

\subsection{AAVGen development}

\subsubsection{Regression models as reward functions}

To enable precise prediction of AAV capsid functional traits, we developed three regression models for each of production fitness, kidney tropism, and thermostability properties. These models were created by fine-tuning an ESM-2 PLM with 8 million parameters. This model architecture was selected for its efficiency in capturing sequence-level representations while maintaining computational feasibility for iterative fine-tuning. We employed a sequential transfer learning strategy to leverage shared functional features across properties: the production fitness model served as the foundational checkpoint, from which the kidney tropism and thermostability models were further adapted. All models were implemented using the Hugging Face Transformers library (version 4.56.1)\cite{Wolf2020}, with \texttt{EsmForSequenceClassification} configured for regression (\texttt{num\_labels=1}). Training involved tokenizing amino acid sequences to a fixed maximum length of 755 residues with padding and truncating as needed. Hyperparameters were optimized for convergence and generalization, incorporating mixed-precision training (FP16), gradient accumulation, and early stopping based on validation loss. Training optimization employed the mean squared error (MSE) as the primary objective function across all three regression tasks with AdamW as optimizer, ensuring consistent gradient behavior and stable convergence. This approach ensured robust, property-specific predictors that could be deployed as reward functions within the GSPO framework.

\paragraph{Production Fitness.}
The regression model for production fitness prediction was trained on the merged AAV2 dataset from Bryant et al. and Ogden et al. studies, comprising deduplicated VP1 sequences with $\log_2$-transformed, WT-normalized fitness scores. After a stratified train-validation split, sequences were tokenized using the ESM-2 tokenizer. Training commenced from the ESM-2, utilizing the AdamW optimizer with a linear learning rate scheduler (initial rate: $1\times10^{-4}$, weight decay: 0.01, warmup steps: 10). We configured 10 epochs, a per-device batch size of 16 (effective batch size 128 via 8 gradient accumulation steps), evaluation every 100 steps, and checkpointing every 50 steps (retaining up to 10 checkpoints).

\paragraph{Kidney Tropism.}
Building upon the production fitness checkpoint (selected at step 14,100), a kidney tropism prediction model was fine-tuned using the processed AAV2 dataset from the Ogden et al. study. This dataset comprised $\log_2$-transformed selection coefficients derived from mouse kidney enrichment ratios, following the exclusion of low- and high-abundance outliers and the aggregation of biological replicates. Tokenized representations of the sequences by the ESM-2 tokenizer were generated before training. Transfer learning was performed using a cosine learning rate scheduler with an initial learning rate of $2\times10^{-6}$, a weight decay of 0.01, and 20 warm-up steps, over a total of 10 epochs. The batch sizes were maintained at 16 for training with the gradient accumulation of 32 (effective batch size of 128) and 32 for evaluation. Model evaluation and checkpointing were conducted every 100 training and evaluation steps, respectively. Early stopping with a patience of three epochs was implemented to mitigate overfitting.

\paragraph{Thermostability.}
The thermostability prediction model was also adapted from the production fitness checkpoint (selected at step 14,100) using the AAV2 dataset from the Ogden et al. study, which included $\log_2$-transformed selection values from temperature-selection assays. Data handling followed the established protocol of aggregation, cleaning, and stratified splitting. Tokenization and dataset formatting were consistent. Due to the subtler signal in thermostability data, we extended training to 500 epochs with a highly conservative cosine scheduler (initial rate: $5\times10^{-7}$, weight decay: 0.01, warmup steps: 50), preserving batch and evaluation settings (100-step intervals, 20-checkpoint limit). Early stopping patience of 20 epochs was used to allow gradual convergence.

These regression models contributed to a multi-objective reward system, quantifying trade-offs in production fitness, kidney tropism, and thermostability to steer generative sampling toward viable AAV variants.

\subsubsection{Supervised fine-tuning of the ProtGPT2}

To enable AAVGen to capture the underlying residue-residue relationships across AAV serotypes, we fine-tuned the pretrained ProtGPT2 PLM using a curated corpus of high-fitness AAV capsid sequences. ProtGPT2 is a decoder-only autoregressive transformer comprising 36 layers with a model dimensionality of 1,280 and approximately 738 million parameters pretrained in a self-supervised manner on the UniRef50 (version 2021\_04)\cite{UniProt2021} protein sequence database with a training set of approximately 44.9 million sequences. It was loaded from the Hugging Face model hub and fine-tuned using the transformer reinforcement learning (TRL) (version 0.23.1)\cite{vonWerra2020} library's \texttt{SFTTrainer} implementation. Training was performed on GPU hardware under mixed-precision (FP16) settings to balance computational speed and numerical stability. The model was trained with a per-device batch size of 4 and a gradient accumulation of 4 steps (effective batch size of 16). Optimization employed the AdamW algorithm ($\beta_1 = 0.9$, $\beta_2 = 0.999$, $\varepsilon = 1 \times 10^{-8}$) with a learning rate of $1 \times 10^{-4}$, following a linear decay schedule incorporating a 1\% warm-up ratio. Weight decay was set to 0.01 to mitigate overfitting, and training was carried out for three epochs with a maximum sequence length of 300 tokens. Model checkpoints were saved every 100 training steps.

\subsubsection{Reinforcement learning with GSPO}

To further optimize the generative model's performance and specifically train it to produce AAV capsid sequences with enhanced thermostability, kidney tropism, and production fitness, we employed a reinforcement learning technique termed GSPO. This approach builds on the principles of policy gradient methods but emphasizes sequence-level rewards, diverging from token-level optimization strategies like Group Relative Policy Optimization (GRPO). GSPO treats the complete generated sequence as the atomic unit for reward computation and policy updates, ensuring that the optimization signal directly reinforces holistic sequence viability. During training, for each sequence $x$ from the dataset $D$, the current policy $\pi_{\theta_{\text{old}}}$ (the model at the start of the training step) generates a group of $G$ candidate sequences:
\begin{equation}
  y_i \sim \pi_{\theta_{\text{old}}}(\cdot \mid x), \quad i = 1, \ldots, G
  \label{eq:1}
\end{equation}
These sequences are evaluated using a suite of reward functions, and the policy parameters $\theta$ are updated to favor high-reward generations while penalizing deviations from the reference policy to maintain stability.

Given a completion $y$ to a sequence $x$, its likelihood under the policy $\pi_\theta$ is denoted as:
\begin{equation}
  \pi_\theta(y \mid x) = \prod_{t=1}^{|y|} \pi_\theta\!\left(y_t \mid x, y_{<t}\right)
  \label{eq:2}
\end{equation}
with $|y|$ denoting the number of tokens in $y$. Each sequence-completion pair $(x, y)$ is scored by a composite reward function $r(x,y)$, which aggregates multiple property-specific signals. The advantage $\hat{A}_i$ for each sequence $y_i$ is estimated via group-wise normalization to reduce variance:
\begin{equation}
  \hat{A}_i = \frac{r(x, y_i) - \bar{r}}{\sigma_r}
  \label{eq:3}
\end{equation}
where $\bar{r} = \frac{1}{G}\sum_{j=1}^{G} r(x, y_j)$ and $\sigma_r = \sqrt{\frac{1}{G}\sum_{j=1}^{G}\left(r(x, y_j) - \bar{r}(x)\right)^2}$.

To quantify policy improvement, GSPO computes a sequence-level importance ratio:
\begin{equation}
  s_i(\theta) = \left(\frac{\pi_\theta(y_i \mid x)}{\pi_{\theta_{\text{old}}}(y_i \mid x)}\right)^{1/|y_i|}
  \label{eq:4}
\end{equation}
which measures the geometric mean deviation of the new policy $\pi_\theta$ from the reference policy $\pi_{\theta_{\text{old}}}$ across the tokens in sequence $y_i$, normalized by sequence length to account for variable generation lengths.

The GSPO loss objective is then formulated as an expectation over the dataset and generations, incorporating a clipped surrogate to prevent destructive large updates:
\begin{equation}
  J_{\text{GSPO}}(\theta) = \mathbb{E}_{\substack{x \sim D,\\ \{y_i\}_{i=1}^G \sim \pi_{\theta_{\text{old}}}(\cdot|x)}}
  \left[\frac{1}{G}\sum_{i=1}^{G} \min\!\left(s_i(\theta)\,\hat{A}_i,\; \mathrm{clip}\!\left(s_i(\theta), 1-\epsilon, 1+\epsilon\right)\hat{A}_i\right)\right]
  \label{eq:5}
\end{equation}
where $\epsilon$ is a clipping hyperparameter (set to 0.2) that bounds the importance ratio within $[1-\epsilon, 1+\epsilon]$. This objective encourages the policy to increase the probability of high-advantage sequences, while clipping mitigates instability from policy shifts. The importance sampling level was configured at the sequence granularity to align with the reward philosophy.

\subsubsection{Reward functions}

To guide the model toward generating AAV capsid sequences with desired properties, we defined a set of sequence-level reward functions. These functions quantify key objectives---production fitness, kidney tropism, and thermostability---alongside auxiliary metrics for sequence length deviation and intra-batch uniqueness. The composite reward $r(x,y)$ was defined as a linear combination of five specialized functions that contributed equally to the multi-objective optimization. Each reward was computed post-generation by first converting from FASTA to normal sequences. Together, they provide a balanced multi-objective signal that promotes both functional optimization and generative diversity.

\paragraph{Production fitness reward.}
Sequences were passed through the production fitness prediction regression model to obtain predicted scores $p_i$. Rewards were mapped via a linear function to emphasize gains over predicted WT production fitness ($w_{\text{fitness}} \approx -2.5$) while accounting for model MAE on the validation set ($e_{\text{fitness}} \approx 0.40$):
\begin{equation}
  r_{\text{fitness}}(y_i) = \begin{cases}
    a & \text{if } p_i > w_{\text{fitness}} + 4e_{\text{fitness}}, \\
    \dfrac{3a}{4} & \text{if } w_{\text{fitness}} + 4e_{\text{fitness}} \geq p_i > w_{\text{fitness}} + e_{\text{fitness}}, \\
    \dfrac{a}{2} & \text{if } w_{\text{fitness}} + e_{\text{fitness}} \geq p_i > w_{\text{fitness}}, \\
    \dfrac{a}{2} & \text{if } w_{\text{fitness}} \geq p_i \geq \dfrac{a}{2}, \\
    p_i & \text{otherwise},
  \end{cases}
\end{equation}
where $a = -\min_j(\tilde{p}_j)$ is the negative minimum of the validation set predictions $(\tilde{p}_j)$. The reward mapping and decision thresholds are shown in Figure~\ref{fig:3}C.

\paragraph{Kidney tropism reward.}
Analogous to fitness, using the kidney tropism prediction model ($w_{\text{kidney}} \approx -0.16$, $e_{\text{kidney}} \approx 0.83$) with the same mapping logic.

\paragraph{Thermostability reward.}
Identical structure via the thermostability prediction model ($w_{\text{thermostability}} \approx -0.43$, $e_{\text{thermostability}} \approx 1.29$).

\paragraph{Sequence length controller reward.}
To ensure that the policy does not converge exclusively toward producing the WT VP1 sequence (with a length $l_{\text{wt}} = 735$), we introduced a sequence length controller reward function with a tolerance parameter $\sigma = 3$. This function encourages exploration of sequence lengths that deviate from the WT while maintaining compact and potentially functional designs. The reward is defined as:
\begin{equation}
  r_{\text{length}}(y_i) = 1 - \exp\!\left(-\frac{\left(l(y_i) - l_{\text{wt}}\right)^2}{2\sigma^2}\right),
\end{equation}
where $l(y_i)$ represents the length of the generated sequence. This formulation penalizes sequences that closely match the WT length (yielding values near zero) and rewards greater deviations, thus promoting diversity in the generated outputs.

\paragraph{Intra-batch uniqueness reward.}
To discourage mode collapse, sequences were checked for duplicates within the group:
\begin{equation}
  r_{\text{unique}}(y_i) = \begin{cases}
    0 & \text{if } \left|\{j : y_j = y_i\}\right| > 1, \\
    1 & \text{otherwise.}
  \end{cases}
\end{equation}

\subsubsection{Implementation and training procedure}

Implementation was carried out using a custom \texttt{GRPOTrainer} extension from the TRL library, adapted for sequence-level surrogates. The supervised fine-tuned ProtGPT2 model served as the initial policy, with sequences constructed as the ProtGPT2 tokenizer's end-of-sequence (EOS) token followed by a ``$\backslash$n'' and the initiating amino acid ``M'' (i.e., $x = \text{EOS} + \text{``}\backslash\text{nM''}$), mimicking the start of a VP1 sequence in a FASTA-like format. For each training step, $G = 32$ sequences were autoregressively sampled per sequence using temperature $= 1.0$, top-p $= 1.0$, and no top-k filtering, with a repetition penalty of 1.0 to encourage diversity. Generations were truncated at a maximum completion length of 754 tokens.

Training spanned 4 epochs, with per-device batch size 4 and gradient accumulation of 8 (effective batch size 32). Optimization used AdamW ($\beta_1 = 0.9$, $\beta_2 = 0.999$, $\varepsilon = 10^{-8}$) at learning rate $2\times10^{-6}$ with cosine decay, weight decay 0.01, max gradient norm 1.0, and 50 warmup steps. FP16 mixed precision, gradient checkpointing, and periodic cache emptying (every 5 steps) ensured memory efficiency. Logging occurred every step, with checkpoints saved per epoch. The initial dataset comprised 500 samples, enabling rapid iteration on the fitness landscape. This GSPO regime refined the policy to produce AAV2-like capsids with superior multi-property profiles, balancing exploration and exploitation through the sequence-centric objective.

\subsection{AAVGen Evaluation}

The performance of AAVGen was evaluated through large-scale inference using a custom generation pipeline. A total of 500,000 protein sequences were generated, each initiated with the fixed token ``M,'' to assess the model's ability to produce coherent outputs. Inference was executed using a batch size of 64, leveraging a sampling-based decoding strategy (temperature $= 1.0$, top-p $= 1.0$, top-k $=$ None) to enable diverse sequence generation, with each sequence limited to a maximum length of 500 tokens. Furthermore, the quality of the generated sequences was evaluated using different modalities, including basic and alignment analysis, prediction of functional properties, and structural modeling.

\subsubsection{Basic and sequence alignment analysis}

To ensure the novelty and integrity of the generated sequences, we applied a multi-step evaluation protocol. To assess the diversity of AAVGen's outputs, we quantified the extent of sequence repetition among the generated proteins. Specifically, we cumulatively sampled non-overlapping subsets of 1,000 sequences from the total pool of 500,000 generated sequences, progressively increasing the sample size until the full set was evaluated. This analysis enabled us to estimate the frequency of repetitive sequences as a function of the number of generated samples. All duplicate sequences were subsequently removed. We then compared the length distribution of the remaining unique generated sequences with that of the training set, as sequences that are excessively long or short may disrupt protein structure and function.

To further quantify the similarity between the generated protein sequences and the WT sequence, we computed a suite of sequence-based metrics. Each generated sequence was globally aligned to the WT sequence using the \texttt{PairwiseAligner} implemented in Biopython (version 1.85)\cite{Cock2009}, with standard scoring parameters (match score $= 2$, mismatch score $= -1$, gap opening penalty $= -2$, gap extension penalty $= -0.5$). From the optimal alignment, we calculated the percentage identity, defined as the proportion of aligned positions with exactly matching amino acids relative to the total alignment length. We also calculated the percentage similarity, defined as the proportion of aligned positions with either identical amino acids or conservative substitutions, as determined by the alignment scoring scheme. In addition to these alignment-based measures, we assessed sequence divergence using the edit distance, defined as the minimum number of single-residue insertions, deletions, or substitutions required to transform the generated sequence into the WT sequence. The distributions of these alignment- and distance-based metrics collectively characterize the relationship between the generated sequences and the WT reference in terms of exact matches, biochemical similarity, and overall divergence.

\subsubsection{Functional property analysis}

To evaluate the generative capacity of AAVGen in predicting production fitness, kidney tropism, and thermostability, we assessed the generated sequences using trained regression models for each functional metric. Predicted scores were obtained for all generated variants and compared to the corresponding AAV2 WT values. Relationships among the predicted functional properties were quantified using Spearman correlation to assess potential trade-offs or co-optimization across metrics.

Based on the reward model design, generated sequences were further stratified according to their predicted performance relative to WT and the uncertainty of the regression models, as estimated by the MAE on the validation set. Variants were classified as ``Best'' if their predicted score exceeded WT by more than four times the MAE, ``Good'' if their score fell between one and four MAE above WT, ``Uncertain'' if their predicted score fell between the WT score and one MAE above it, and ``Bad'' if their predicted score was lower than that of WT.

This functional stratification enabled the selection of representative variants across a range of predicted score levels for downstream structural analysis. By integrating functional predictions with structural modeling, we aimed to determine whether improvements or degradations in predicted properties were associated with measurable changes in capsid structure.

\subsubsection{Structural alignment and analysis}

Protein structures were generated using AlphaFold3, which is a diffusion-based structure prediction deep learning model. To assess the structural similarity of generated sequences with respect to the WT protein, we used the PDB structure of AAV2 obtained from the Protein Data Bank (PDB)\cite{wwPDB2019}, which represents only the VP3 subunit responsible for forming the AAV capsid surface. From 500,000 generated sequences, after the preprocessing phase and removal of all the sequences labeled as ``Bad'' and ``Uncertain'', we randomly sampled 500 sequences for structural prediction. To establish a baseline for structural comparison, we first aligned the 500 sampled sequences with the WT sequence and identified a region of variability where specific residues were mutated in at least 20 sequences. Within this region, we randomly inserted amino acids, taking into account the length distribution observed in the 500 generated sequences. This process resulted in a dataset of 250 randomly generated sequences. For each sequence, we sampled 5 structures to ensure robustness. Structural alignment between the predicted mutant and WT structures was performed in PyMOL (version 3.1.1)\cite{PyMOL2015}. The RMSD was then calculated as the square root of the average squared distance between equivalent C$\alpha$ atoms after optimal superposition, reflecting the mean positional deviation between the two structures. Additionally, C$\alpha$ atoms with positional differences within 0.5~\AA{} were selected, expanded to their corresponding residues, and residues with larger deviations were identified as differing regions. Finally, we analyzed the distribution of RMSD and evaluated the correlation between predicted scores and RMSD.

\subsection{Hardware and training time}

The computational resources for model training consisted of a dedicated server featuring an NVIDIA V100 GPU with 32~GB of VRAM and an AMD Epyc 7502 CPU with 32~GB of RAM. Under this configuration, the training durations for the models were: 11 hours and 25 minutes for the fitness regression model; 3 hours and 24 minutes for the kidney regression model; 3 hours and 29 minutes for the thermostability regression model; 9 hours and 5 minutes for the SFT phase of training model; and 9 hours and 38 minutes for the GSPO phase.

% -----------------------------------------------------------------------
\newpage
\section*{Data availability}

The data used in this study are publicly available on Hugging Face: the input dataset can be accessed at \href{https://huggingface.co/datasets/Moreza009/AAV_datasets}{Moreza009/AAV\_datasets} and the generated output dataset produced in this study is available at \href{https://huggingface.co/datasets/Moreza009/AAVGen-dataset-out}{Moreza009/AAVGen-dataset-out}

\section*{Code availability}

All code for this study is publicly available on GitHub (\href{https://github.com/mohammad-gh009/AAVGen}{mohammad-gh009/AAVGen}). The AAVGen model and associated regression models for production fitness, kidney tropism, and thermostability are available on Hugging Face at: \href{https://huggingface.co/Moreza009/AAVGen}{Moreza009/AAVGen}, \href{https://huggingface.co/Moreza009/AAV-Fitness}{Moreza009/AAV-Fitness}, \href{https://huggingface.co/Moreza009/AAV-Kidney-Tropism}{Moreza009/AAV-Kidney-Tropism}, and \href{https://huggingface.co/Moreza009/AAV-Thermostability}{Moreza009/AAV-Thermostability}, respectively. An inference notebook for AAVGen is additionally available on Kaggle (\href{https://www.kaggle.com/code/mohammadgh009/aavgen}{notebook}).

\section*{Acknowledgment}

Not applicable.

\section*{Funding}

No funding was received for this study or its publication.

\section*{Competing interest}

The authors declare no competing interests.

\section*{Authors contribution}

\textbf{Conceptualization:} M.G, Y.G. \textbf{Dataset preparation:} M.G. \textbf{Model development:} M.G. \textbf{Model assessment:} M.G, Y.G. \textbf{Data interpretation:} M.G, Y.G. \textbf{Drafting original manuscript:} M.G, Y.G. \textbf{Revising the manuscript:} M.G, Y.G. All the authors have read and approved the final version for publication and agreed to be responsible for the integrity of the study.

% -----------------------------------------------------------------------
\clearpage
\bibliographystyle{unsrtnat}

\end{document}